\documentclass[a4paper]{PoS}

\usepackage{amssymb, fontenc, times, mathptmx, graphicx}
\usepackage{amsfonts}
\usepackage{mathtools}
\usepackage{amsmath}
\usepackage{mathrsfs}
\usepackage{subcaption}
\usepackage{tikz-feynman}
\tikzfeynmanset{compat=1.1.0}
\usepackage{xfrac}
\usepackage{array}% For table's line thickness change
\usepackage{bm}

\title{$SU(7)$ Unification of Scotogenic Model with Pati-Salam $SU(4)_c\otimes SU(2)_L\otimes U(1)_R$ gauge symmetry}

\ShortTitle{Pati-Salam+Scotogenic model+$SU(7)$}

\author{Arnab Dasgupta$^a$, Sin Kyu Kang$^a$ and \speaker{Oleg Popov}$^b$\\
		\llap{$^a$}School of Liberal Arts, Seoul-Tech, Seoul 139-743, Korea\\
		\llap{$^b$}Institute of Convergence Fundamental Studies, Seoul National University of Science and Technology, Seoul 139-743, Korea\\
		E-mail:  \email{skkang@seoultech.ac.kr}, \email{arnabdasgupta28@gmail.com},
        \email{opopo001@ucr.edu}}

\abstract{In this work we demonstrate how Scotogenic model can arise naturally from an $SU(7)$ grand unified theory. Our model also includes a low scale Pati-Salam $SU(4)_c\times SU(2)_L\times U(1)_R$ symmetry which unifies quarks and leptons. $\mathbb{Z}_2$ symmetry needed for the realization of Scotogenic scenario is a residual symmetry and not \emph{ad hoc}. A short fermion masses and dark matter analysis is included.}

\FullConference{The 39th International Conference on High Energy Physics (ICHEP2018)\\
		4-11 July, 2018\\
		Seoul, Korea}


\begin{document}

\section{Introduction}
\label{sec:into}
Standard Model (SM) has been very successful in explaining many aspects of particles and their interactions but it predictions massless neutrinos. In order to generate neutrino masses one needs to go beyond the SM. There exist numerous proposals for generating Majorana (generated through the dim-5 Weinberg operator~\cite{Weinberg:1979sa}, $\frac{\text{LLHH}}{\Lambda}$) as well as Dirac neutrino masses. The minimal cases are Seesaw-I (SM+(1,1,0)$_F$), Seesaw-2 (SM+(1,3,1)$_S$), and Seesaw-3 (SM+(1,3,0)$_F$), which are tree level Majorana cases. At 1-loop there are Zee model~\cite{Zee:1980ai}, Scotogenic model~\cite{Ma:2006km}, and other less known Scotogenic inverse seesaw realization~\cite{Fraser:2014yha}. All possible pathways to tree and one-loop radiative Majorana and Dirac neutrino mass generation have been studied in ref.~\cite{Ma:1998dn} and ref.~\cite{Ma:2016mwh} respectively. It shows that there are 4 possible ways to generate Dirac neutrino masses at tree level and only two possible topologies exist for one-loop Dirac neutrino mass.

The aim of this letter is to demonstrate how Scotogenic model\cite{Ma:2006km} of neutrino masses, which relates the existence of dark matter to the generation of neutrino masses, can be naturally obtained from the $SU(7)$ grand unified theory (GUT) breaking with quarks and leptons unified through the Pati-Salam $SU(4)_c\times SU(2)_L\times U(1)_R$ gauge symmetry at the low scale ($\sim O(10^3)$TeV).

In reference~\cite{Ma:2014kph} it was shown how the $\mathcal{Z}_2$ and $U(1)_D$ Scotogenic models can be naturally obtained from $SU(6)$ and $SU(7)$ GUT breaking respectively but without additional intermediate symmetries at low energy scales.

The note is organized as follows: description of the model, unification of gauge couplings, fermion mass analysis, and short discussion of dark matter candidates and analysis.

\section{Model}
\label{sec:model}

Our model consists of the $\mathbb{G}_{\text{GUT}}=SU(7)$ with $\underline{\textbf{21}}_F, \underline{\textbf{7}}_F^*, \underline{\textbf{35}}_F^*$ fermion and $\underline{\textbf{140}}_S, \underline{\textbf{28}}_S$ scalar fields. The model is chiral anomaly free, $\frac{3}{2}-\frac{1}{2}-1=0$. The $SU(7)$ GUT is then spontaneously broken to $\mathbb{G}_{PS}=SU(4)_c\times SU(2)_L\times U(1)_R$ by $\underline{\textbf{140}}_S$. The Yukawas of the model are $\underline{21}_F\times\underline{35}_F^*\times\underline{140}_S$, $\underline{21}_F\times\underline{7}_F^*\times\underline{140}_S^*$, $\underline{35}_F\times\underline{35}_F\times\underline{140}_S$, $\underline{7}_F^*\times\underline{7}_F^*\times\underline{28}_S$ which generate the masses for all fermions. $\underline{\textbf{28}}_S$ scalar field is required here in order to generate the Majorana mass for the neutral $\mathbb{Z}_2$ odd fermion in the Scotogenic model. Next, $\mathbb{G}_{PS}$ is broken down to $\mathbb{G}_{SM}=SU(3)_c\times SU(2)_L\times U(1)_Y$ at $\sim O(10^3$TeV$)$ by $\underline{\textbf{140}}_S$. Electric charge, hypercharge, electroweak isospin, and $U(1)_R$ charge are $\scriptstyle{Q=\left(I_3+Y\right)_{SM}=\left(I_3+R+\frac{\sqrt{6}}{3}T_4\right)_{PS}=\left(\sqrt{\frac{2}{3}}T_3+\sqrt{\frac{2}{5}}T_4+\frac{7}{\sqrt{15}}T_5+\sqrt{\frac{7}{3}}T_6\right)_{SU(7)}, Y=\left(R+\frac{\sqrt{6}}{3}T_4\right)_{PS}, \left(I_3\right)_{SM,PS}=\left(-\sqrt{\frac{2}{5}}T_4+\sqrt{\frac{3}{5}}T_5\right)_{SU(7)}}$,\\ $\scriptstyle{R=\left(\frac{4}{\sqrt{10}}T_4+\frac{4}{\sqrt{15}}T_5+\sqrt{\frac{7}{3}}T_6\right)_{SU(7)}}.$
The model also includes one diagonal generator at the GUT scale, here labeled as $Z_7$ which is $\scriptstyle{Z_7=\left(\frac{\sqrt{7}}{5}T_4+\sqrt{\frac{14}{3}}\frac{1}{5}T_5-2\sqrt{\frac{2}{15}}T_6\right)_{SU(7)}}.$
List of all vacuum expectation values (VEV) for this model is given in Tab.~\ref{tab:vev1}.
\small{
\begin{table}
\centering
\begin{tabular}{ccccc}
\# & Scale & Rep$_{SU(7)}$ & Rep$_{PS}$ & Rep$_{SM}$ \\
1 & PS & $\underline{28}_S$ & $\left(10,1,1\right)$ & $\left(1,1,0\right)$ \\
2 & EW & $\underline{28}_S$ & $\left(4,2,0\right)$ & $\left(1,2,-1/2\right)$ \\
3 & EW & $\underline{28}_S$ & $\left(1,3,-1\right)$ & $\left(1,3,-1\right)$ \\
4 & EW & $\underline{140}_S$ & $\left(1,3,-1\right)$ & $\left(1,3,-1\right)$ \\
5 & EW & $\underline{140}_S$ & $\left(15,2,-1/2\right)$ & $\left(1,2,-1/2\right)$ \\
6,7 & EW & $\underline{140}_S$ & $\left(1,2,-1/2\right)$ & $\left(1,2,-1/2\right)$ \\
8 & GUT & $\underline{140}_S$ & $\left(1,1,0\right)$ & $\left(1,1,0\right)$ \\
9 & EW & $\underline{140}_S$ & $\left(4,2,0\right)$ & $\left(1,2,-1/2\right)$ \\
10 & EW & $\underline{140}_S$ & $\left(4,3,1/2\right)$ & $\left(1,3,0\right)$ \\
11,12 & PS & $\underline{140}_S$ & $\left(4,1,1/2\right)$ & $\left(1,1,0\right)$ \\
13 & EW & $\underline{140}_S$ & $\left(4,2,1\right)$ & $\left(1,2,1/2\right)$
\end{tabular}
\caption{List of vacuum expectation values for the $SU(7)_{GUT}$ model.}
\label{tab:vev1}
\end{table}
}
Decomposition of $SU(7)$ $\underline{\textbf{21}}_F, \underline{\textbf{7}}_F^*, \underline{\textbf{35}}_F^*$ fermion and $\underline{\textbf{140}}_S, \underline{\textbf{28}}_S$ scalar fields under $SU(7)\rightarrow SU(4)_c\times SU(2)_L\times U(1)_R\rightarrow SU(3)_c\times SU(2)_L\times U(1)_Y$ breaking is given in Ref.~\cite{Kang:2018ab}.

\section{Neutrino Masses}
\label{sec:mnu}

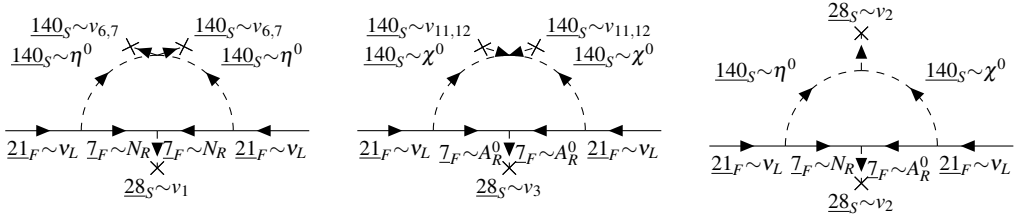
\begin{figure}[h]
\centering
\begin{subfigure}{0.3\textwidth}
\begin{tikzpicture}
\begin{feynman}
\vertex (i);
\vertex [right =1cm of i] (a);
\vertex [right =1cm of a] (b);
\vertex [below =0.5cm of b] (bb) {$\scriptstyle{\underline{28}_S\sim v_1}$};
\vertex [right =1cm of b] (c);
\vertex [right =1cm of c] (f);
\vertex [above =1cm of b] (tc);
\vertex [above =0.354cm of tc] (x1);
\vertex [left =0.354cm of x1] (tl) {$\scriptstyle{\underline{140}_S\sim v_{6,7}}$};
\vertex [right =0.354cm of x1] (tr) {$\scriptstyle{\underline{140}_S\sim v_{6,7}}$};
\diagram*[small]{
(i) -- [fermion, edge label' = $\scriptstyle{\underline{21}_F\sim\nu_L}$] (a) -- [fermion, edge label' = $\scriptstyle{\underline{7}_F\sim N_{R}}$] (b) -- [anti fermion, edge label' = $\scriptstyle{\underline{7}_F\sim N_{R}}$] (c) -- [anti fermion, edge label' = $\scriptstyle{\underline{21}_F\sim\nu_L}$] (f),
(a) -- [charged scalar, quarter left, edge label = $\scriptstyle{\underline{140}_S\sim\eta^{0}}$] (tc)[cross],
(c) -- [charged scalar, quarter right, edge label' = $\scriptstyle{\underline{140}_S\sim\eta^{0}}$] (tc)[cross],
(tc) -- [charged scalar, insertion = 0.99] (tl),
(tc) -- [charged scalar, insertion = 0.99] (tr),
(b) -- [charged scalar, insertion = 0.99] (bb),
};
\end{feynman}
\end{tikzpicture}
%\caption{Scotogenic neutrino mass loop mediated by $N\sim (\textbf{1},\textbf{1},0)_{SM}$.}
\label{fig:mnuN}
\end{subfigure}
\begin{subfigure}{0.3\textwidth}
\begin{tikzpicture}
\begin{feynman}
\vertex (i);
\vertex [right =1cm of i] (a);
\vertex [right =1cm of a] (b);
\vertex [below =0.5cm of b] (bb) {$\scriptstyle{\underline{28}_S\sim v_3}$};
\vertex [right =1cm of b] (c);
\vertex [right =1cm of c] (f);
\vertex [above =1cm of b] (tc);
\vertex [above =0.354cm of tc] (x1);
\vertex [left =0.354cm of x1] (tl) {$\scriptstyle{\underline{140}_S\sim v_{11,12}}$};
\vertex [right =0.354cm of x1] (tr) {$\scriptstyle{\underline{140}_S\sim v_{11,12}}$};
\diagram*[small]{
(i) -- [fermion, edge label' = $\scriptstyle{\underline{21}_F\sim\nu_L}$] (a) -- [fermion, edge label' = $\scriptstyle{\underline{7}_F\sim A^0_{R}}$] (b) -- [anti fermion, edge label' = $\scriptstyle{\underline{7}_F\sim A^0_{R}}$] (c) -- [anti fermion, edge label' = $\scriptstyle{\underline{21}_F\sim \nu_L}$] (f),
(a) -- [charged scalar, quarter left, edge label = $\scriptstyle{\underline{140}_S\sim\chi^{0}}$] (tc)[cross],
(c) -- [charged scalar, quarter right, edge label' = $\scriptstyle{\underline{140}_S\sim\chi^{0}}$] (tc)[cross],
(tc) -- [anti charged scalar, insertion = 0.99] (tl),
(tc) -- [anti charged scalar, insertion = 0.99] (tr),
(b) -- [charged scalar, insertion = 0.99] (bb),
};
\end{feynman}
\end{tikzpicture}
%\caption{Scotogenic neutrino mass loop mediated by $A^0\in (\textbf{1},\textbf{2},-\frac{1}{2})_{SM}$.}
\label{fig:mnuA}
\end{subfigure}
\begin{subfigure}{0.3\textwidth}
\begin{tikzpicture}
\begin{feynman}
\vertex (i);
\vertex [right =1cm of i] (a);
\vertex [right =1cm of a] (b);
\vertex [below =0.5cm of b] (bb) {$\scriptstyle{\underline{28}_S\sim v_2}$};
\vertex [right =1cm of b] (c);
\vertex [right =1cm of c] (f);
\vertex [above =1cm of b] (tc);
\vertex [above =0.5cm of tc] (tt){$\scriptstyle{\underline{28}_S\sim v_{2}}$};
\diagram*[small]{
(i) -- [fermion, edge label' = $\scriptstyle{\underline{21}_F\sim\nu_L}$] (a) -- [fermion, edge label' = $\scriptstyle{\underline{7}_F\sim N_{R}}$] (b) -- [anti fermion, edge label' = $\scriptstyle{\underline{7}_F\sim A^0_{R}}$] (c) -- [anti fermion, edge label' = $\scriptstyle{\underline{21}_F\sim\nu_L}$] (f),
(a) -- [charged scalar, quarter left, edge label = $\scriptstyle{\underline{140}_S\sim\eta^{0}}$] (tc)[cross],
(c) -- [charged scalar, quarter right, edge label' = $\scriptstyle{\underline{140}_S\sim\chi^{0}}$] (tc)[cross],
(tc) -- [charged scalar, insertion = 0.99] (tt),
(b) -- [charged scalar, insertion = 0.99] (bb),
};
\end{feynman}
\end{tikzpicture}
%\caption{Scotogenic neutrino mass loop mediated by $N\sim (\textbf{1},\textbf{1},0)_{SM}$ and $A^0\in (\textbf{1},\textbf{2},-\frac{1}{2})_{SM}$.}
\label{fig:mnuNA}
\end{subfigure}
\caption{Scotogenic neutrino mass loop mediated by $N\sim (\textbf{1},\textbf{1},0)_{SM}$(left), $A^0\in (\textbf{1},\textbf{2},-\frac{1}{2})_{SM}$(center), and mixing of $N\sim (\textbf{1},\textbf{1},0)_{SM}$ and $A^0\in (\textbf{1},\textbf{2},-\frac{1}{2})_{SM}$(right).}
\label{fig:mnu1}
\end{figure}

Neutrino masses are generated radiatively through the $\mathbb{Z}_2$ odd sector, just like in Scotogenic model. Since our model has 3 neutral colorless fermions, there will be 2 neutral $\mathbb{Z}_2$ odd fermions to mediate the radiative neutrino mass diagram. In order to get scotogenic scenario to work $v_5$ and $v_{10}$ VEV's from Tab.~\ref{tab:vev1} need to be 0. The corresponding radiative neutrino mass diagrams are depicted in Fig.~\ref{fig:mnu1}. Here $N\sim(\bm{1},\bm{1},0)_{SM}$ and $A\sim(\bm{1},\bm{2},-\frac{1}{2})_{SM}$.

\section{Unification}
\label{sec:uni}

Unification is achieved at $M_{GUT}\sim O(10^{12-16}\text{TeV})$ scale with intermediate $SU(4)_c\times SU(2)_L\times U(1)_R$ Pati-Salam symmetry at $\sim O(10^3\text{TeV})$. We used the following boundary conditions during gauge coupling unification: {$\alpha_s(m_Z)=0.1182$, $\alpha_{em}(m_Z)=1/127.916$, $\alpha_2(m_Z)=\frac{\sqrt{2}m_W^2 G_F}{\pi}=0.03393$. The intermediate Pati-Salam symmetry breaking scale is taken to be $M_{PS}=10^3$ TeV as in~\cite{Perez:2013osa}, and U(1) normalization given by $n_Y=\sqrt{\frac{17}{3}}$ and $n_R=\sqrt{5}$. Then $\alpha=\frac{g^2}{4\pi}$ at unification scale is $\alpha_U(M_U)=0.08577$ with $M_U=10^{12}$ GeV. This model, due to large amount of scalar fields, can easily predict the weak angle at $m_Z$ scale to be given as Sin$^2\theta_W(m_Z)=\frac{3}{8}$. The plot showing gauge coupling unification is given in Fig.~\ref{fig:uni}. When achieving gauge coupling unification we took only the relevant fields, fields from Ref.~\cite{Perez:2013osa} and fields necessary for radiative neutrino mass generation, to be at the Pati-Salam scale or lighter with all other fields at the $M_U$ GUT scale.

\begin{figure}
\centering
\includegraphics[scale=0.5,trim={1cm 21cm 1cm 2cm}, clip]{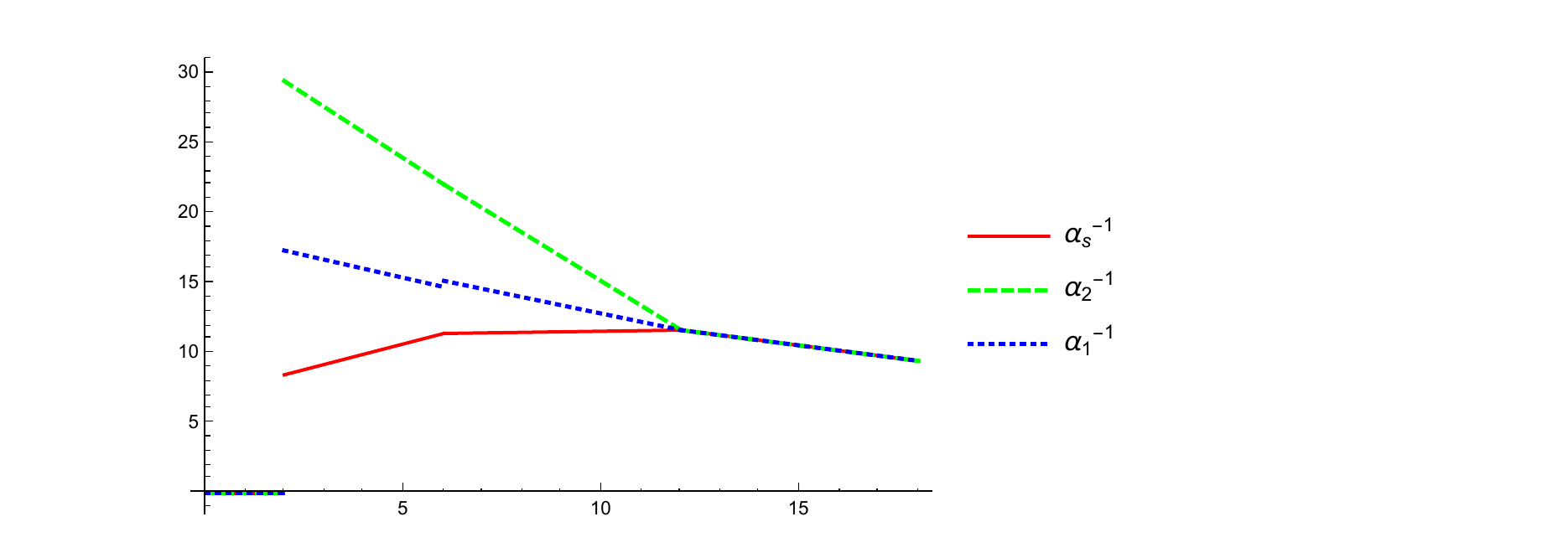}
\caption{$SU(7)$ gauge coupling unification at $M_{GUT}\sim O(10^{12-16}\text{TeV})$ with intermediate $SU(4)_c\times SU(2)_L\times U(1)_R$ Pati-Salam symmetry at $\sim O(10^3\text{TeV})$. Horizontal axis indicates log scale $\mu$ in units of GeV.}
\label{fig:uni}
\end{figure}

\section{Fermion Masses}
\label{sec:ferm_m}

Fermion masses for our model are given as \\
$\scriptstyle{\scriptstyle{
\left(\begin{matrix}
v_{5,7} & v_8 & v_{12} & v_{13} \\
v_{10} & v_{9} & v_{7} & v_{4} \\
v_{13} & v_{11} & v_{8} & v_{6} \\
v_{4} & v_{5,6} & v_{9} & v_{10,11}
\end{matrix}\right)_{M_{\pm}},
\left(\begin{matrix}
    v_{10,11} & v_5 \\
    v_{6,7} & v_{11,12}
    \end{matrix}\right)_{M_{\pm2}},
\left(\begin{matrix}
0 & v_5 & v_{10} \\
v_5 & v_1 & v_2 \\
v_{10} & v_2 & v_3
\end{matrix}\right)_{M_{Q=0}},
\left(\begin{matrix}
    v_5 & 0 \\
    v_{10} & v_{5}
    \end{matrix}\right)_{M_{\pm 2/3}},
\left(\begin{matrix}
v_{5,7} & v_{8} & v_9 & v_{11,12} \\
v_{10} & v_{13} & v_{4} & v_{5,6,7} \\
v_{9} & v_{11} & v_{5,6,7} & v_{8} \\
v_{4} & v_{5,6} & v_{10,12} & v_{13}
\end{matrix}\right)_{M_{\pm 1/3}}
}},$\\
$\scriptstyle{\scriptstyle{\left(\begin{matrix}
   0 & v_{5} \\
   v_5 & v_{10,12}
   \end{matrix}\right)_{M_{\pm 4/3}}
   }}.$

$v_{i}$ entries in the mass matrices indicate to what VEV that entry is proportional to. Our model includes 4 Dirac $Q=\pm 1$ leptons, 4 down-type quarks, 2 up-type quarks, 3 neutral fermions, 2 exotic Dirac $Q=\pm 2$ leptons, 2 exotic up-type $Q=\pm 4/3$ quarks. Exotic particles can be used to test this model. Focusing on neutral mass matrix we can see that for the scotogenic scenario we need $v_5$ and $v_{10}$ to be 0 which will split neutral fermion mass matrix in $\mathbb{Z}_2$ even and odd parts. This requirement will also predict the up-type $Q=\pm 2/3$ quark masses to be generated radiatively as well, due to quark-lepton unification symmetry. Another interesting thing to notice is that, because there are 3 neutral fermions predicted by the model we can naturally generate inverse see-saw, linear see-saw, double see-saw, \emph{etc.}~\cite{Ma:2009du} in this model if we give up scotogenic scenario.

\section{Dark Matter}
\label{sec:dm}
Any of the $\mathbb{Z}_2$ odd particles can be considered for light stable particle and be the dark matter candidate. A brief analysis of dark matter relic vs dark matter mass is given in Fig.~\ref{fig:dm}. Scalar, pseudo-scalar, fermion dark matter candidates are possible in our model.

\begin{figure}
\centering
\begin{subfigure}{0.3\textwidth}
\includegraphics[width=\textwidth]{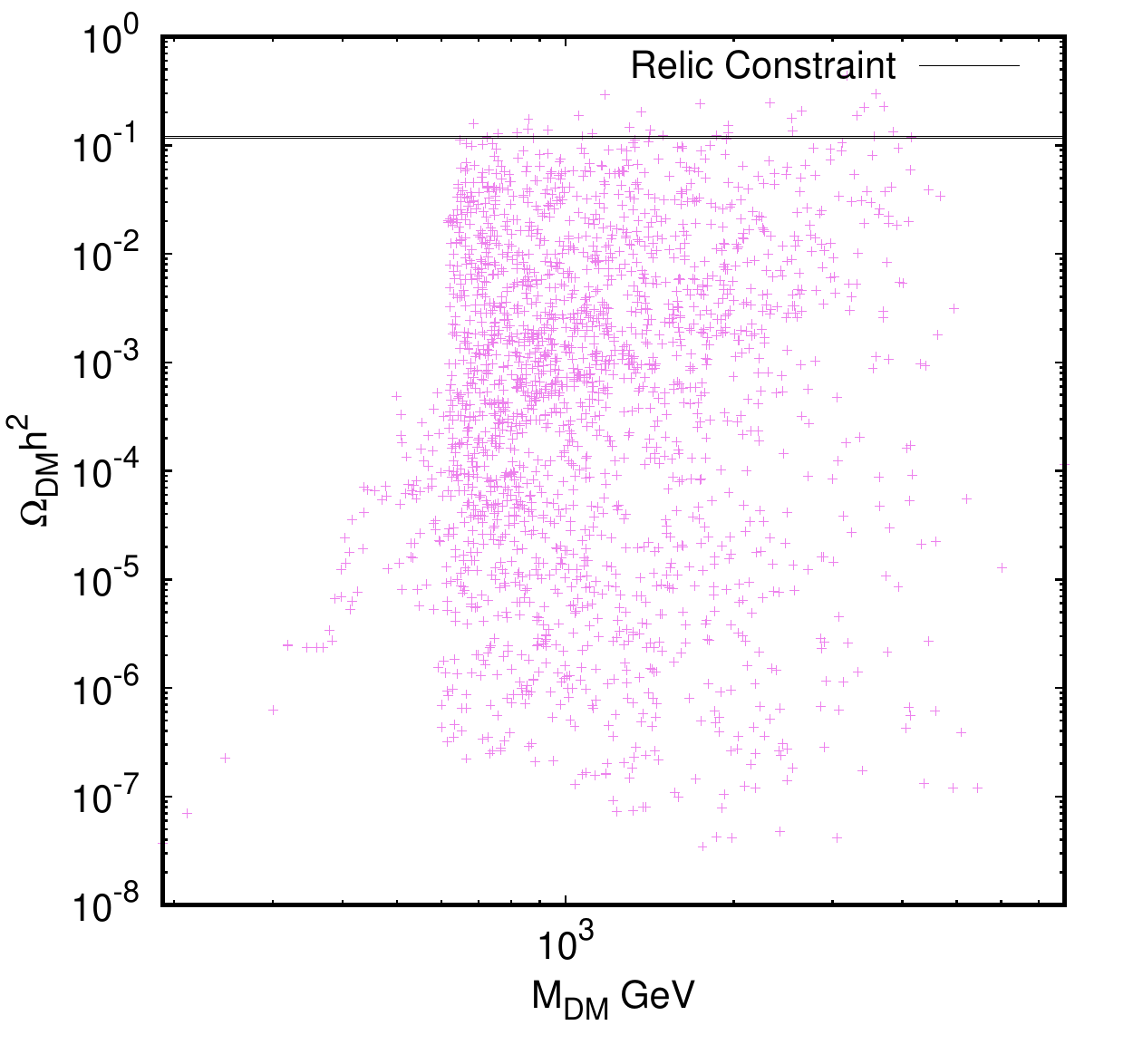}
\end{subfigure}
\begin{subfigure}{0.3\textwidth}
\includegraphics[width=\textwidth]{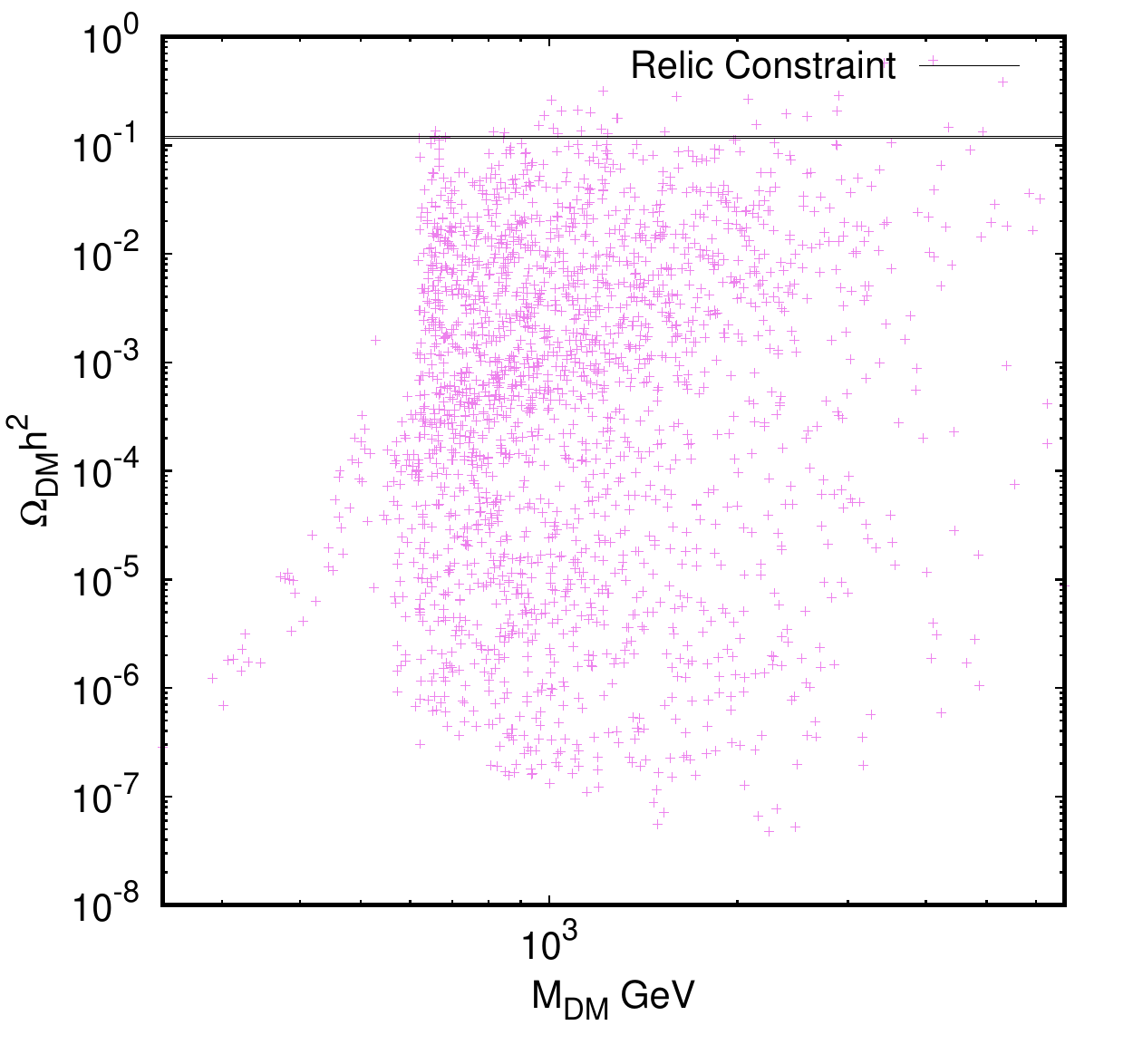}
\end{subfigure}
\begin{subfigure}{0.3\textwidth}
\includegraphics[width=\textwidth]{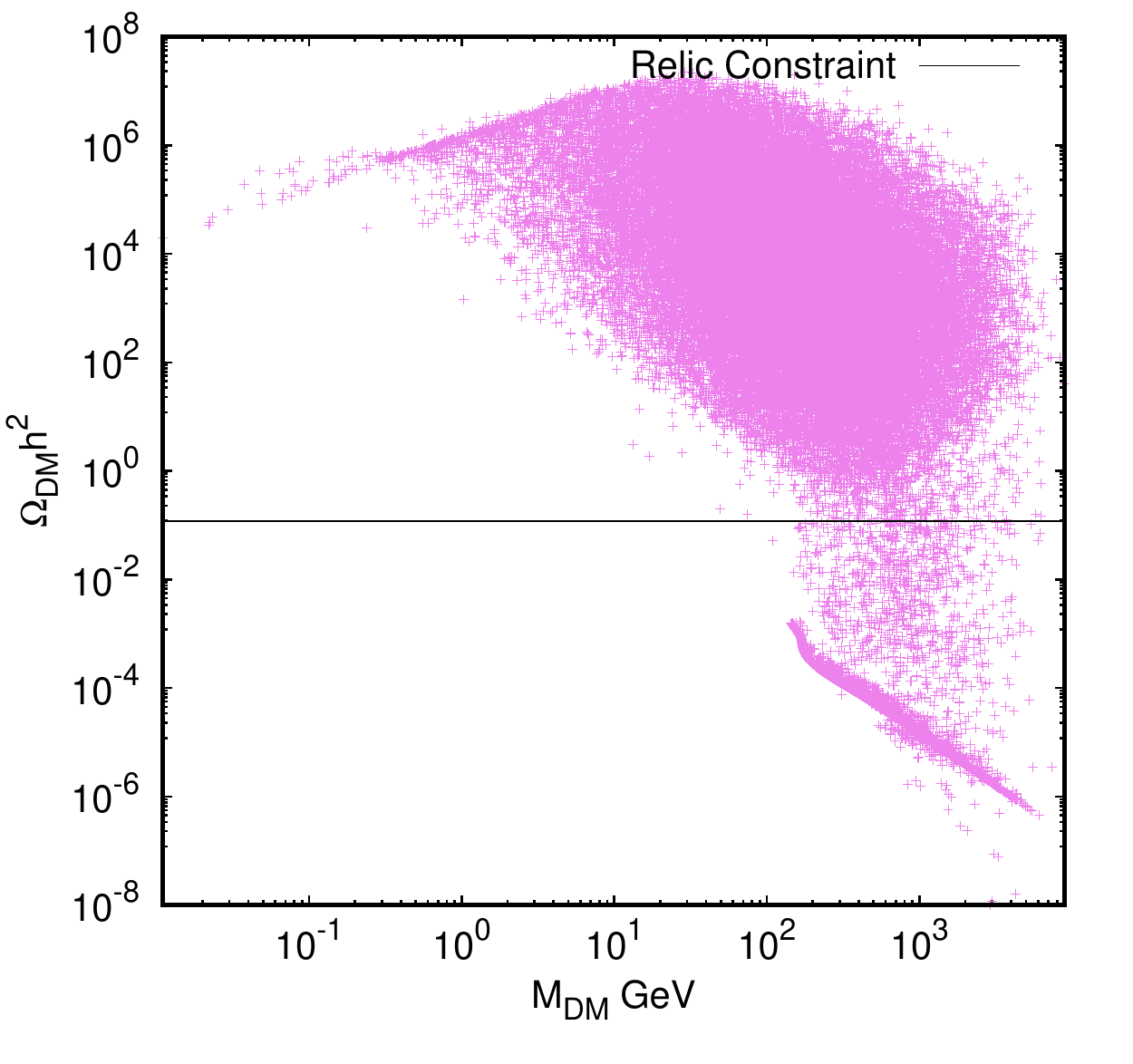}
\end{subfigure}
\caption{Dark matter relic density vs dark matter mass for real scalar DM (left), pseudo-real scalar DM (center), and fermionic DM (right).}
\label{fig:dm}
\end{figure}

\section{Conclution}
\label{sec:conc}

In conclusion we have shown how Scotogenic model with residual $\mathbb{Z}_2$ symmetry can be obtained from $SU(7)$ GUT with low scale ($O(\sim 10^3\text{TeV})$) quark-lepton unification through Pati-Salam symmetry. We have shown how neutrino masses are radiatively generated at one-loop, gauge couplings are unified at GUT scale $~O(10^{12-16}\text{GeV})$ with correct weak angle prediction at electroweak scale. We have also shown how this model can naturally accommodate inverse seesaw, linear seesaw, double seesaw, \emph{etc} scenarios. Brief discussion of dark matter candidates with analysis is included.

\end{document}